\documentclass[twocolumn,showpacs,preprintnumbers,amsmath,amssymb,floatfix]{revtex4}

\usepackage{graphicx}

\newcommand{\Sigt}{\Sigma_t}
\newcommand{\vc}[1]{{#1}^\alpha}   
\newcommand{\Der}{\delta} 
\newcommand{\thl}{\theta_l}
\newcommand{\thk}{\theta_k}
\newcommand{\eps}{\epsilon} 

\newcommand{\half}{\frac{1}{2}}

\newcommand{\la}{\langle} 
\newcommand{\ra}{\rangle} 

\usepackage{amsthm}
\newtheorem{theorem}{Theorem}
\newtheorem{definition}{Definition} 
\newtheorem{proposition}{Proposition}
\newtheorem{lemma}{Lemma}

\begin{document}

\title{Local existence of dynamical and trapping horizons}
 
\author{Lars Andersson} \email{larsa@math.miami.edu} 
\thanks{Supported in part by the NSF,
contract no. DMS 0104402.}
\affiliation{Albert
Einstein Institute, Am M\"uhlenberg 1, D-14476 Potsdam, Germany}
\affiliation{Department of Mathematics, University of Miami, Coral Gables, FL
33124, USA}
\author{Marc Mars} \email{marc@usal.es} 
\thanks{Supported in part by the 
Spanish Ministerio de Educaci\'on y Ciencia, project 
BFME2003-02121.} 
\author{Walter Simon}
\email{walter@usal.es} 
\thanks{Supported in part by the Austrian FWF, project no. P14621-N05.} 
\affiliation{Facultad de Ciencias, Universidad de
Salamanca\\ Plaza de la Merced s/n, E-37008 Salamanca, Spain.}

\begin{abstract}
Given a spacelike foliation of a spacetime and a marginally outer trapped surface S on some initial leaf, 
we prove that under a suitable stability condition S is contained in a ``horizon'', i.e. a smooth 3-surface 
foliated by marginally outer trapped slices which lie in the leaves of the given foliation.  
We also show that under rather weak energy conditions this horizon must be either achronal or spacelike 
everywhere. Furthermore, we discuss the 
relation between ``bounding'' and ``stability'' properties of marginally outer trapped surfaces.
\end{abstract}

\maketitle

The application of numerical relativity to black hole spacetimes
is, together with the role played by black hole thermodynamics as a testing
ground for quantum gravity, among the factors that have
caused a shift of interest from global properties of black holes such as 
the event horizon, knowledge of which requires information about the infinite
future, towards quasilocal properties. By quasilocal properties one means  
such properties
that can at least in principle be measured by an observer with a finite life
span and hence also can be studied during the course of a numerical
evolution of a black hole spacetime.

A closed spacelike surface $S$ in a spacetime $(M,g_{\alpha\beta})$ is 
called trapped if future directed null rays emanating from $S$ are
converging. If  $M$ contains a trapped surface,
satisfies the null energy condition and some causality condition,
then $M$ is singular \cite{RP}.
Suppose $M$ is foliated by a family of spacelike Cauchy surfaces $\{\Sigt\}$. 
The apparent horizon, defined as the family of boundaries of the
regions containing trapped surfaces in the $\{\Sigt\}$, is a 
quasilocally defined object that plays an important
role in black hole thermodynamics, as well as in numerical evolutions of
black holes. It should be noted however that the apparent horizon
depends on the choice of the reference foliation $\{\Sigt\}$.
If sufficiently smooth, the apparent horizon is foliated by marginally outer 
trapped surfaces (MOTS) \cite{KH}. The latter are
 defined to have vanishing outgoing null expansion, (while 
the  ingoing one is not restricted).

In numerical evolution of black hole spacetimes, it is now standard to 
avoid the singular behavior of both gravitational field and gauge
conditions in the interior of black holes, by excising a suitable region
inside the boundary of the black holes, as defined by a collection of MOTS,
from the computational domain. However, tracking a family of MOTS during an
evolution one encounters the occasional ``sudden'' appearance of
new MOTS and by ``jumps'' of the MOTS (see, e.g. \cite{BH}). 
It thus becomes important to study such an evolution analytically as far as
possible. In this Letter we prove, in Theorem 1, existence of a horizon, 
i.e. a hypersurface foliated by MOTS, provided the initial surface $S$ satisfies 
a natural stability condition, and in Theorem 2 we give causal properties of $H$.
The  condition of ``strictly stably outermost'',  which is crucial for these
results, means that there is an outward deformation of $S$ such that the corresponding
variation of the outgoing null expansion is nonnegative and positive somewhere
(c.f. Definition \ref{def:stablyouter} for details). 
\begin{theorem} \label{thm1} 
Let $(M, g_{\alpha\beta})$ be a smooth spacetime foliated by smooth spacelike hypersurfaces $\Sigt$. 
Assume that some leaf $\Sigma = \Sigma_0$ contains a smooth marginally outer trapped surface $S$ 
which is strictly stably outermost.  

Then, $S$ is contained in a smooth horizon $H$ whose marginally outer trapped leaves lie in $\Sigt$, 
and which exists at least as long as these marginally trapped leaves remain strictly stably
outermost.
\end{theorem}
In Theorem 2 below we use the same notation as in Theorem 1, 
and we denote by $\vc{l}$ the null vector for which the expansion 
$\theta_l$ vanishes on $S$. Recall that the null energy condition
holds if $G_{\alpha\beta} j^\alpha j^\beta \geq 0$ for any null vector $\vc{j}$, where 
$G_{\alpha\beta}$ is the Einstein tensor. 

\begin{theorem}
\label{thm2}
If, in addition to the hypotheses of Theorem 1, the null energy condition holds, 
the horizon $H$ is locally achronal. 
If, moreover,  $G_{\alpha\beta} l^\alpha l^\beta > 0$ somewhere
on $S$ or if $S$ has nonvanishing shear with respect to $\vc{l}$ somewhere, 
then  $H$ is spacelike everywhere near $\Sigma$. \\
\end{theorem}
\vspace*{-3mm}
The term {\it horizon} in this Letter is closely related to {\em dynamical horizons}
introduced by  Ashtekar and Krishnan \cite{AK} and to {\em trapping horizons} 
introduced by Hayward \cite{SH}. 
The latter two are more restrictive in the sense that they also require  
$\thk < 0$ for the expansion along the other future null direction $k$. 
While dynamical horizons are spacelike by definition, trapping horizons may have any causal
character a priori, but they are required to satisfy 
an additional stability condition, namely that the variation of $\thl$ along 
$k$ is negative. Our condition of ``strictly stably outermost'' can be generalized to variations
in the outward past null cone $C_{-}$ of $S$. Hayward's stability
condition is then closely related to our stability condition along null directions.
Most of the results stated in this Letter, in particular the existence theorem, extend
to the null case. Details will be given elsewhere \cite{AMS}. 
%


Since the location in spacetime  of apparent, dynamical and trapping horizons 
depends in general on the foliation $\Sigt$, it is clear that the same applies 
to the horizons obtained above.
However, for generic dynamical horizons, this dependence is limited by 
the uniqueness results of Ashtekar and Galloway \cite{AG}.
We also note that the result in Theorem 2 on the causal character of $H$ 
is stronger than the one known \cite{SH} for trapping horizons.

\begin{figure}[!ht]
\centering \includegraphics[width=0.45\textwidth]{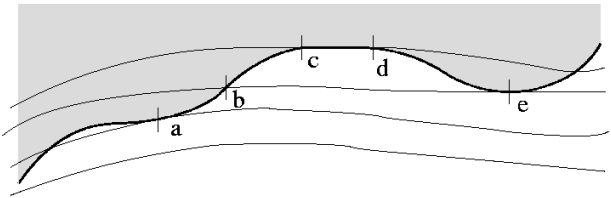}
\caption{A horizon}
\label{fig:horizon}  
\end{figure} 
The example illustrated in Fig. 1 will shed light on the possible behavior of
our horizons.  It shows a  horizon (the thick line) in a spacetime
foliated by spacelike  hypersurfaces $\Sigt$ (thin lines). 
The horizon separates a region in which the $\Sigma_t$ contain outer trapped surfaces 
(which we call ``trapped region'', shaded in Fig. \ref{fig:horizon}) 
from a region where the $\Sigma_t$ are free of them; the
intersection ``points'' with  the foliation are MOTS. 
Note that Fig. \ref{fig:horizon}  incorporates naturally the observed sudden appearance of
MOTS during the evolution (e.g. point e).  If the numerical
analysis only looks for globally outermost MOTS (as it is usually done) it is
clear  that they  jump (e.g., from b to e), while they are in fact connected by a horizon 
interpolating between both which  ``runs downward'' in some places (in the interval (d,e)). 
Thus Fig. \ref{fig:horizon} makes compatible a smooth horizon with the jumps observed
numerically. Examples like Fig. \ref{fig:horizon} can, in particular, 
be constructed in spherically symmetric spacetimes by choosing the spacelike foliation $\Sigt$
suitably. Thus one may expect that the situation described by Fig. \ref{fig:horizon}  
is typical, though the causal character of the "downward" part of the horizon 
is yet unclear; see the discussion in \cite{AG}.

We now introduce some notation needed for the precise statements of our
results. All fields and manifolds will be assumed to be $C^\infty$
unless otherwise stated.  
 Let $(M, g_{\alpha\beta})$ be a spacetime with signature $-+++$.
Given a spacelike surface $S$ in $M$ we may choose two future
directed null fields $\vc{l}, \vc{k}$.
Recall that the {\bf variation} $\delta_p \nu$ 
of the geometric object $\nu$ defined on a surface $T$ in the direction of 
the vector $p^{\alpha}$ is defined by $\delta_p \nu = \partial \nu/\partial \tau$ 
for any one-parameter family of surfaces $T_{\tau}$ with $T_{0}=T$ and
$p^{\alpha}\partial_{x^{\alpha}} = \partial/\partial \tau |_{\tau=0}$. 
The null expansion $\thl$ is  defined by $\mu \thl = \Der_{l} \mu$, where
$\mu$ is the volume form on $S$. It should be noted 
that the variation is additive in the sense that 
for example  $\Der_{\psi k + l} \thl = \Der_{\psi k} \thl +  \Der_l
\thl$, for some function $\psi$, 
but in general $\Der_{\psi k} \thl \ne \psi \Der_k \thl$. 

A closed spacelike surface $S$ is called outer trapped (weakly outer trapped, 
marginally outer trapped) if one of the null expansions,  say $\thl$, is negative
(non-positive, zero) everywhere on $S$.
 (For an alternative terminology, c.f. \cite{MS}). The usual definition of
``trapped surface'' and ``marginally trapped surface'' requires additional
conditions on the expansion with respect to the  other null vector $\vc{k}$. 
Corresponding definitions for untrapped surfaces are made by reversing the 
signs. Let a reference foliation $\{\Sigt\}$ of $M$ by spacelike
hypersurfaces be given, and select one such surface $\Sigma =
\Sigma_0$.  For a MOTS $S \subset \Sigma$,
we define the ``outward'' direction within $\Sigma$ as the one to which the
projection to $\Sigma$ of the null vector $\vc{l}$ selected above points. This
definition of ``outward'' need not coincide with the intuitive one in
asymptotically flat spacetimes. However, all our results hold for arbitrary
spacetimes (not necessarily asymptotically flat) and $\vc{l}$ defines a local
concept of ``outward'' for MOTS. The  unit outward normal
to $S$ tangent to $\Sigt$ is called $\vc{m}$, the future pointing unit normal to
$\Sigt$ is $\vc{n}$, and we scale the null vectors $\vc{l}$ and $\vc{k}$ such that $\vc{l} = \vc{n}
+ \vc{m}$ and  $\vc{k} = \vc{n} - \vc{m}$.

The following definitions are, apart from later use, motivated by similar
definitions of  Newman \cite{RN} and of  Kriele and Hayward \cite{KH}, 
and by results in these papers.
\begin{definition} A marginally outer trapped surface $S$ is called
{\bf locally outermost} in $\Sigma$, iff there exists a two-sided
neighbourhood of $S$ such that its exterior part does not contain any
 weakly outer trapped surface.
\end{definition} 
\begin{definition}\label{def:stablyouter} 
A marginally outer trapped surface  $S$ is called {\bf stably outermost} iff
there exists a function $\psi \geq 0$, $\psi \not\equiv 0$, on $S$
such that  $\Der_{\psi m} \thl \geq 0$.  $S$ is called {\bf strictly stably
outermost} if, moreover, $\Der_{\psi m} \thl \ne 0$  somewhere on $S$.
\end{definition} 
In Fig. \ref{fig:horizon}, the points in the interval [a,d]
represent stably outermost surfaces, those in [a,c) locally outermost ones, and
those in (a,c)  strictly stably outermost ones. 
This example, and the result in  \cite{KH} suggest the implications:
Strictly stably outermost $\Rightarrow$  Locally outermost $\Rightarrow$
Stably outermost, and the picture also suggests counterexamples  for the
opposite directions. We now give the tools required to show these
results and the Theorems.

For a function $\psi$ on $S$, we define a linear elliptic
operator  $L_\Sigma$ by $L_\Sigma \psi = \Der_{\psi m} \thl$.
Explicitly, we obtain 
\begin{align}
\label{lop}
L_\Sigma \psi &= 
 - \Delta_{S} \psi + 2 s^A D_A \psi +  \left( \frac{1}{2} R_{ S} - 
 s_A s^A + \right. \nonumber  \\
 &
+  \left. D_A s^A - \frac{1}{2} K_{AB}^{\mu} K^{\nu \, AB}l_{\mu}l_{\nu}
- G_{\alpha\beta} l^{\alpha} n^{\beta} \right) \psi.
\end{align} 
Here $D_A$ is the covariant derivative on $S$, $\Delta_{S}$
is the corresponding Laplacian, $R_S$ is the scalar curvature,
$K^\mu_{AB}$  is the second fundamental form vector (defined by 
$K^\mu_{AB} v_\mu = - \nabla_A v_B$ for any normal $v_\alpha$ to $S$,
where $\nabla_\alpha$ is the covariant derivative on $(M,g_{\alpha\beta})$)
 and $s_A$ is the torsion of $\vc{l}$ (the 1-form 
$ s_A = -\half k_\alpha \nabla_A l^\alpha $ on $S$).

$L_\Sigma$ is analogous to the stability operator for minimal surfaces.
In general, $L_\Sigma$ is not self-adjoint but the eigenvalues of $L_\Sigma$ have
their real part bounded
from below. The eigenvalue with smallest real part is called the {\bf principal eigenvalue}. 
The following holds for second order elliptic operators of the form of $L_\Sigma$. 
\begin{lemma}
The principal eigenvalue $\lambda$ of $L_\Sigma$ is real. 
Moreover, the corresponding principal eigenfunction $\phi$  
(which satisfies $L_\Sigma \phi= \lambda \phi$) 
is either everywhere positive or everywhere negative.
\end{lemma}

This Lemma is a consequence of the 
Krein-Rutman theorem which can be applied to 
second order elliptic operators along the lines 
in corollary A3 of Smoller \cite{JS}. The discussion in 
Smoller's corollary can be adapted straightforwardly to 
the case without boundary. 

We now restate Definition \ref{def:stablyouter} in terms of $\lambda$
as follows.
\begin{lemma} \label{lem:eigen} 
Let $S \subset \Sigma$ be a MOTS and  
let $\lambda$ be the principal eigenvalue
of the corresponding operator $L_{\Sigma}$. Then $S$ is stably outermost iff
$\lambda \geq 0$ and strictly stably outermost iff $\lambda >0$.
\end{lemma}
\begin{proof} 
If $\lambda \geq 0$, choose $\psi$ in the definition of (strictly) stably
outermost as a positive eigenfunction $\phi$ corresponding to $\lambda$.
Then $\Der_{\phi m} \thl =  L_{\Sigma} \phi = \lambda \phi \geq 0$.
For the converse, we note that the adjoint $L^{\ast}_\Sigma$  
(with respect to the standard $L^2$ inner
product $\la \, , \ra$ on $S$) has the same principal
eigenvalue as $L_\Sigma$, and a positive principal eigenfunction
$\phi^{\ast}$. 
Thus, for $\psi$ as in the definition of (strictly)
stably outermost,
$$
\lambda \la \phi^{\ast}, \psi \ra =
\la L_{\Sigma}^{\ast} \phi^{\ast}, \psi \ra = \la \phi^{\ast},
L_{\Sigma} \psi  \ra \geq 0,
$$
with strict inequality in the strictly stable case. Since $\la
\phi^{\ast}, \psi \ra > 0$, the Lemma follows.
\end{proof} 
\begin{proposition}  \mbox{}
\label{Implication1}
\begin{itemize}
\item[(i)] 
A strictly stably outermost surface $S$  is locally outermost.
Moreover, $S$ has a two-sided neighbourhood $U$ such
that no weakly outer trapped surfaces contained in $U$ enter the exterior of $S$ and
no weakly outer untrapped surfaces contained in $U$ enter the interior of $S$.
\item[(ii)] A locally outermost surface $S$ is stably outermost.
\end{itemize}
\end{proposition}
\begin{proof} 
The first statement of (i) is in fact contained in the second one. 
To show the latter, let $\phi$ be the positive principal eigenfunction of $L_\Sigma$. Since
$L_\Sigma \phi > 0$ by assumption, flowing $S$ in $\Sigma$ 
along any extension of $\phi m^{\alpha}$ produces a family
$S_\sigma$, $\sigma \in (-\eps,\eps)$ for some $\eps > 0$. By choosing $\eps$
small enough, the $S_\sigma$ have $\thl |_{S_\sigma} > 0$ 
for $\sigma \in (-\eps, 0)$ and $\thl |_{S_\sigma} < 0$ for 
$\sigma \in (0,\eps)$. We can now take $U$ to be the
neighbourhood of $S$ given by $U = \cup_{\sigma \in (-\eps,\eps)} S_\sigma$. 

Now let $B$ be a weakly outer trapped surface contained in $U$ which enters the exterior part of $U$. 
Then the function $\sigma$ has a maximum $\sigma_p > 0$ at some point $p$ in $B$. 
At $p$, $B$ is tangent to $S_{\sigma_p}$ and we have 
\begin{equation}
\label{eq}
\thl |_B \leq 0 < \thl
|_{S_{\tau_p}} .
\end{equation}  
For a surface represented as a graph with respect to a function $f$, the map
$f \to \thl$ is a quasilinear elliptic operator and 
the strong maximum principle \cite{GT} applies to show that  
the inequality (\ref{eq}) holds only if $B$ coincides with $S_{\tau_p}$
and hence $\thl |_B = \thl |_{S_{\tau_p}}$ which gives a
contradiction. Hence $B$ cannot enter the exterior part of $U$. 

To show (ii), assume $S$ is locally outermost but not stably outermost. From Lemma 2, the principal 
eigenvalue $\lambda$ is then negative. Arguing as above one constructs a foliation outside $S$ with
leaves which are outer trapped near $S$, contradicting the
assumption.
\end{proof} 
Theorem 5.1 of Ashtekar and Galloway \cite{AG} implies that the domain 
exterior to $S$ to which outer trapped surfaces cannot enter is determined 
by the past domain of dependence of any dynamical
horizon through $S$,  provided that some genericity conditions hold.
Outer trapped surfaces ``far outside'' of a locally outermost MOTS
might exist in general (as Fig. \ref{fig:horizon} 
suggests for the surface in the interval (b,c)).
To exclude this, one could define ``globally outermost'' surfaces
(in particular in an asymptotically flat context).  
We can now prove our main theorem.

{\em Proof of Theorem 1.}
Consider a foliation $\Sigt$ with a
MOTS $S$ on $\Sigma = \Sigma_0$.  Let $C_+$ be the null cone generated
by null rays starting from $S$ in the direction of $l^{\alpha}$ and let $\tilde S_t
= C_+ \cap \Sigt$ for $t$ close to $0$. We now introduce coordinates 
$(t, r, x^A)$ in a neighborhood of $S$ such that at $S$, $\partial_r$ 
is the normal $m^{\alpha}$ and $\partial_t$ is parallel to $l^{\alpha}$.

On $\Sigt$ we consider
surfaces which are given as graphs $r= f(x^A)$ in this coordinate system and
we define a functional $\Theta[t,f]$ whose value is $\thl$ on the surface,
and which acts on $f$ as a quasilinear elliptic operator of the form
$
\Theta[t,f] = a^{AB} (f, \partial f) \partial_A \partial_B f + b
(f,\partial f)
$
where the coefficients $a^{AB}$ and  $b$ are smooth functions depending on
$x$ and on $t$, $f$ and
$\partial_A f$.  For integer $k \geq 0$, $\alpha \in (0,1)$, let 
$C^{k,\alpha}$ be H\"older spaces on $S$. 
Let $I = (-\eps,\eps)$ for $\eps > 0$. 
One checks that for some $\eps > 0$, and for any $k \geq 2$, there are
neighborhoods $\mathcal U_1$ and $\mathcal U_2$ of zero in $C^{k,\alpha}$ and
$C^{k-2,\alpha}$ respectively so that  
$\Theta[t, f]: I \times \mathcal U_1 \rightarrow 
\mathcal U_2$ is a well-defined $C^\infty$ map. 
Let $D_y \Theta$ be the derivative with respect to the second
argument. Then we have from the definition of $\Theta$, 
$
D_y \Theta [0,0]. \beta  = L_\Sigma \beta .
$
As $S$ is assumed to be strictly stably
outermost, the principal eigenvalue of $L_\Sigma$ is positive by Lemma
\ref{lem:eigen}. By the Fredholm alternative $L_\Sigma$ is invertible as a
map $L_\Sigma : C^{k,\alpha} \to C^{k-2,\alpha}$ for arbitrary $k \geq 2$,
$\alpha \in (0,1)$. Now the implicit function theorem for Banach space maps
\cite{SL} applies to prove local existence of a smooth horizon $H$ which by 
construction has the property that the leaves $S_t = H \cap \Sigt$ near $S$ are 
MOTS. By patching charts together it is also clear that existence holds as long 
as the $S_t$ stay strictly stably outermost. $\hfill \Box$

\vspace{3mm} 

A theorem of Schoen \cite{RS} asserts the existence of a MOTS between
barrier surfaces $S_1, S_2$ with $S_1$ trapped and $S_2$ untrapped
if the dominant energy condition holds.
 Besides its clear interest, this result also suggests an alternative
approach to existence of a horizon: 
Start from a weakly outer trapped surface $S$ on some initial slice and
take the null cone emanating from it.  By the Raychaudhuri equation, 
 $\Der_{\omega l} \thl = - \omega W$ for any function $\omega$ and
 $W = K_{AB}^\mu K^{\nu AB} l_\mu l_\nu + G_{\mu\nu} l^\mu l^\nu $.
If $W > 0$, the  null cone  cuts each subsequent slice on an outer trapped surface
which gives the trapped barrier $S_1$, while an untrapped barrier $S_2$ 
always exists near infinity for asymptotically flat data. 
Schoen's result then yields existence of a MOTS on every subsequent slice. 
The resulting ``horizon'' may in general jump (e.g. from b to e in Fig.
\ref{fig:horizon}), but it need not be outermost. 
On the other hand, our Theorem 1 requires a MOTS on the initial  surface 
instead of just a trapped one, but we do not assume any asymptotic properties, 
and we obtain a smooth horizon.

Results on the causal character of the horizon can be obtained  by combining again 
the Raychaudhuri equation on the null cone as sketched above with a maximum principle 
or a ``barrier argument'' inside $\Sigma_t$ for $t$ near $\Sigma_0$.  
For the latter, we may use e.g. part (i) of Proposition 1.
However, we can also do ``both steps at once'' by using the following Lemma.
\begin{lemma} \label{lem:ext}
For a strictly stably outermost surface $S$, any normal variation
$\psi \vc{m}$ of $S$ with $\Der_{\psi m} \thl \ge 0$ satisfies  $\psi \geq
0$, i.e. the variation cannot be directed to the interior anywhere on $S$.
If, moreover, $\Der_{\psi m} \thl \ne 0$ somewhere then $\psi > 0$,
i.e. the variation is directed to the exterior everywhere on $S$.
\end{lemma}
\begin{proof}
Let $\phi$ be a positive principal eigenfunction of $L_\Sigma$ and define
$\chi$ by $\psi = \chi \phi$. 
A computation shows that
$$
L_\Sigma \psi =  \chi L_\Sigma \phi - \phi \Delta_{S} \chi + 
2 \left (\phi s^A - D^A \phi \right) D_A \chi.
$$
Since $S$ is strictly stably outermost, 
$L_\Sigma \phi  > 0$  and the strong maximum principle 
\cite{GT} yields that $\chi \geq 0$ if
$L_\Sigma \psi \geq 0$, with strict inequality if 
$L_\Sigma \psi \neq 0$ somewhere.
\end{proof}
We can now prove Theorem 2.\\
{\em Proof  of Theorem 2.} 
By construction, the variation of $\thl$ vanishes along the  vector $q^\alpha
\partial_{x^\alpha} =  \partial_t  + \mu
\partial_r = \omega l  + \mu m$ tangent to $H$, where $\omega > 0$ is defined by $\partial_t = \omega l$. 
Therefore,
$$
0 = \Der_q \thl = \Der_{\omega l + \mu m} \thl = -\omega W + L_\Sigma \mu .
$$
If the null energy condition holds, $W$ is non-negative and under the condition
in the second part of Theorem 2, $W$ is even positive somewhere. 
By Lemma \ref{lem:ext} it follows that $\mu \ge 0$
in the first case and  $\mu > 0$ everywhere in the second one, 
which proves the assertions. $\hfill \Box$ \\ 

Assume that a 2-surface $S$ is strictly stably outermost and that the dominant energy condition 
is satisfied (i.e. that $-G^{\alpha}_{\beta}u^{\beta}$ is future directed 
for all future directed timelike vectors $u^{\alpha}$).
Then $S$ is topologically $S^2$ (and hence the horizon through $S$ is
$S^2 \times \Re$). We recall here Newman's proof \cite{RN} of this fact.
Denoting by  $\phi$ the positive principal eigenfunction of
$L_\Sigma$, the stability condition implies that $0 < \int_S \phi^{-1} L_\Sigma \phi$. 
Then the result follows from (\ref{lop}) after integrating by parts and using the 
Gauss-Bonnet theorem.

While we have restricted ourselves to local results in this Letter, it would
clearly be desirable to determine the global evolution of the horizon
in a given black hole spacetime.
It would be much more ambitious to look at the global Cauchy
evolution for asymptotically flat initial data with a MOTS.
In vacuum, one expects this evolution to approach a Kerr spacetime and our horizon 
to approach the event horizon. Moreover, the area of the  marginally trapped slices 
should approach a quantity not greater than $16\pi m^2$ where $m$ 
is the mass of the final Kerr black
hole. This version of the Penrose inequality 
\cite{BHI} would involve the area of the MOTS 
instead of a minimal surface and one could, for axially symmetric data, include angular
momentum as well \cite{AK}. 
\vspace*{-5mm}
\begin{acknowledgments}
We wish to thank Greg Galloway, Thomas Hoffmann-Ostenhof, Jos\'e Senovilla and Helmuth Urbantke
for helpful discussions and correspondence.
\end{acknowledgments}



\end{document}